\documentclass[%
 twocolumn, aps,
 jmp,%
 amsmath,amssymb,
]{revtex4}
\usepackage{bm}
\usepackage[colorlinks=true,linkcolor=blue]{hyperref}
\usepackage{amsmath}
\usepackage{amssymb}
\usepackage{amsthm}
\usepackage{amsfonts}
\usepackage{times}
\usepackage{enumerate}
\usepackage{latexsym}
\usepackage{ifpdf}
\usepackage{graphicx}
\usepackage{makeidx}
\expandafter\ifx\csname package@font\endcsname\relax\else
 \expandafter\expandafter
 \expandafter\usepackage
 \expandafter\expandafter
 \expandafter{\csname package@font\endcsname}%
 \fi
\usepackage{color}

\begin{document}

\title{Effect of polar discontinuity on the growth of LaNiO$_{3}$/LaAlO$_{3}$ superlattices}
\author{Jian Liu$^1$, M. Kareev$^1$, S. Prosandeev$^1$, B. Gray$^1$, P. Ryan$^2$, J. W. Freeland$^2$ and J. Chakhalian$^1$}

\affiliation{$^1$Department of Physics, University of Arkansas, Fayetteville, AR 72701}
\affiliation{$^2$Advanced Photon Source, Argonne National Laboratory, Argonne, IL 60439}

\begin{abstract}
We have conducted a detailed microscopic  investigation of  [LaNiO$_{3}$(1 u.c.)/LaAlO$_{3}$(1 u.c.)]$_N$ superlattices grown on (001) SrTiO$_{3}$ and LaAlO$_3$ to explore the influence of polar mismatch on the resulting electronic and structural  properties. Our data demonstrate that the initial growth on the non-polar SrTiO$_{3}$ surface leads to a rough morphology and unusual 2+ valence  of Ni in the first LaNiO$_3$ layer, which is not observed after growth on the polar surface of LaAlO$_3$.  A newly devised model suggests that the polar mismatch can be resolved if  the perovskite  layers grow with an excess of LaO, which also  accounts for the observed electronic, chemical, and structural effects.
\end{abstract}

\maketitle

Recently complex oxide ultra-thin films have been drawing enormous attention  due to the possibility of stabilizing unusual quantum phases and building interface-controlled devices \cite{ahn06,Caviglia}. Towards this goal, the problem of polar interfaces  is of fundamental importance, since a polar crystal structure grown on a non-polar substrate would be unstable due to the depolarizing fields \cite{Ohtomo, Noguera}. To avoid this issue, some mechanism must set in to compensate the potential jump by forcing electronic, structural or chemical (non-stoichiometry) changes. In particular, the active work on the  LaAlO$_3$/SrTiO$_3$ system \cite{hwang2,Hwang3, huijbenrev} has suggested an electron transfer between atomic layers that decreases Ti valence by 0.5 at the expense of 0.5$e$ charge in the terminating atomic plane. Additionally, an alloy effect has been observed  at the interface \cite{Muller}, and first-principles computations revealed a contribution of ionic displacements \cite{Pentcheva}.

In this letter, we report on the microscopic observation  of an unusual valence state of Ni$^{2+}$ in ultrathin LaNiO$_{3}$/LaAlO$_{3}$ superlattices (SL) grown on non-polar (or weakly polar) (001) TiO$_{2}$-terminated SrTiO$_3$ (STO) substrates. This is in marked contrast to identical SLs grown on polar LaAlO$_3$ (LAO), which always exhibit the 3+ valence of the bulk-like LaNiO$_3$ (LNO). To clarify the origin of these phenomena, we have tracked the changes of the Ni valence as a function of the SL thickness and developed an experimental procedure to monitor and control the electronic state of a transition metal ion.


High-quality  epitaxial LNO/LAO SLs were grown on (001) TiO$_{2}$-terminated STO single crystal substrates by pulsed laser deposition with \textit{in situ} monitoring by RHEED  \cite{Kareev1}. To minimize induced  defects, STO substrates were prepared by our recently developed chemical wet-etch procedure (`Arkansas treatment') \cite{Kareev2}. A complementary set of   SLs were also grown on LAO single crystal substrates.  Detailed spectroscopic information was acquired in the soft x-ray regime  in both fluorescence yield (FY) mode and total electron yield (TEY) mode at  the Ni L$_{3,2}$ absorption edge at the 4ID-C beamline of the Advanced Photon Source, ANL. To obtain precise information on the Ni charge state, all spectra were aligned by simultaneously measuring a NiO (Ni$^{2+}$) standard with the SLs.  Synchrotron-based x-ray diffraction has confirmed the high structural quality and the full epitaxy of the SLs \cite{Kareev1}.

In order to maintain  the morphological quality and achieve  the layer-by-layer (LBL) growth of the single unit-cell thin  SLs, we used a recently developed interrupted growth method\cite{Kareev1, Blank}. This method consists of  a  rapid ablation  and a prolonged delay between two successive unit cell layers. The usual oscillation of RHEED specular intensity (RSI) for each unit cell occurs within a short flux-on period seen as a sharp dip of RSI, followed by a further slow recovery (see inset in Fig.\ 1). While RSI would have a full recovery in the ideal LBL growth, this first layer grown on STO always exhibits a characteristic drop as seen in Fig. 1, regardless of whether the initial layer is LAO or LNO.  This observation is also consistent with the rough initial growth of the polar layer on the non-polar STO surface ($i.e.$ polar mismatch) as  corroborated  by AFM imaging of the surface. Conversely, this phenomenon  is absent during the initial growth on LAO,  where a full recovery can be seen after the first deposited layer.
\begin{figure}[t]\vspace{-0pt}
\includegraphics[width=8.5cm]{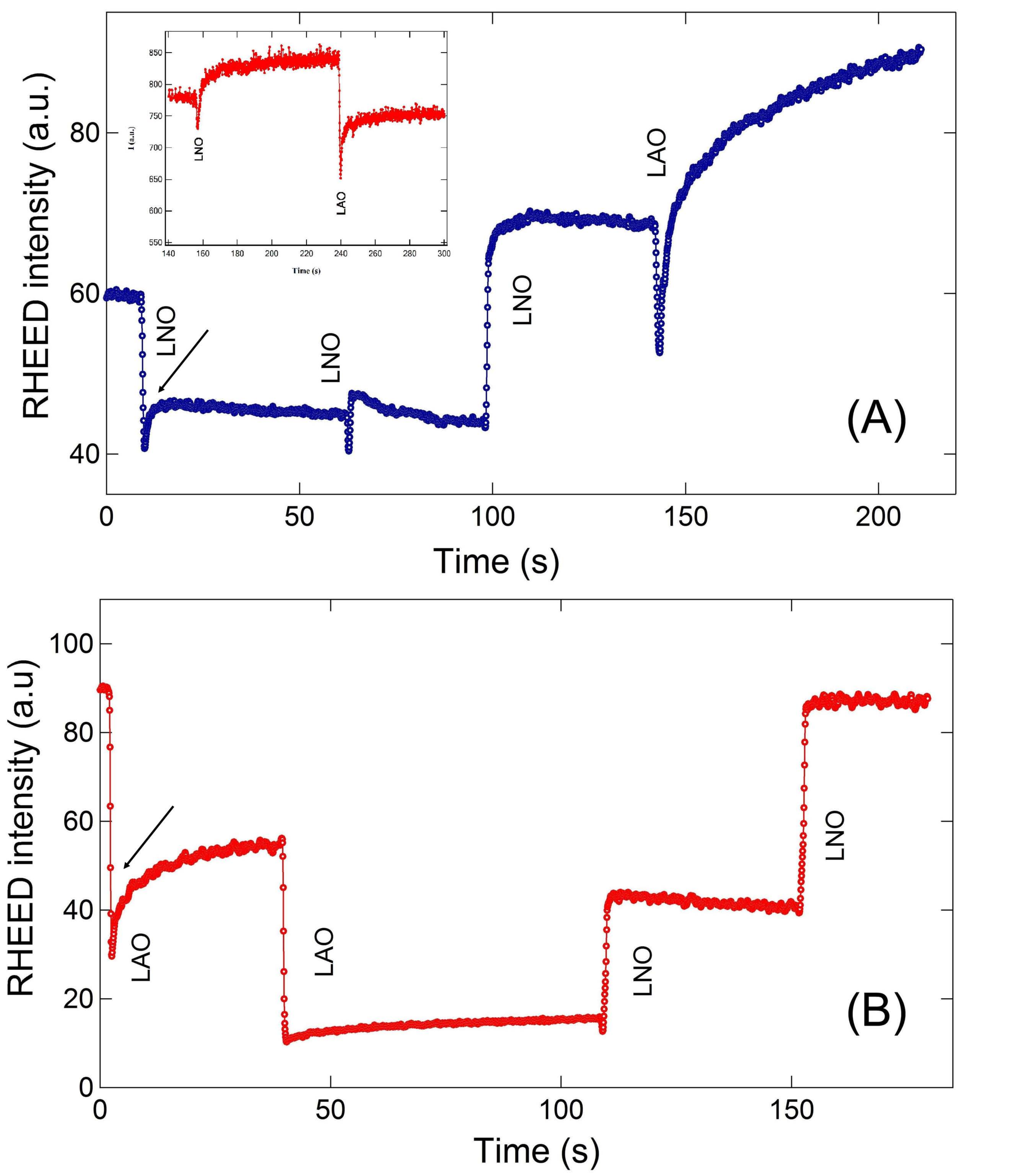}
\caption{\label{RHEED} RHEED specular intensity of the initial growth (marked by black arrows) of LNO (A) and LAO (B) layers on a non-polar STO substrate. Inset of A: Growth of a single repeat of (LNO/LAO) on the polar LAO substrate.}
\end{figure}

To elucidate the microscopic origin of this phenomenon, we performed  resonant soft x-ray spectroscopy (XAS)  on the  Ni L$_{3,2}$-edge.   By using XAS, the Ni valence state can be directly fingerprinted  for samples as thin as a single unit cell. By utilizing this exquisite  sensitivity  to the Ni charge state, we were able to detect and trace the effects of  charge transfer, local symmetry and non-stoichiometry across the layers.  Due to the overlapping La M$_4$-edge, the spectrum at the Ni L$_3$-edge is strongly distorted which is the reason for focusing on the Ni L$_2$-edge. Fig. 2 shows the  Ni L$_2$ spectra measured in the bulk sensitive FY mode for the  [LaNiO$_{3}$(1 u.c.)/LaAlO$_{3}$(1 u.c.)]$_N$ (SLs ([1/1]$_{N}$ thereafter) with a varying number of repeats, \textit{N}, and different growth sequences. A thicker SL composed of  5 u.c. thick LNO layers is also included as a Ni$^{3+}$ reference for a stoichiometric sample in the metallic phase\cite{Piamonteze}. The bulk like Ni$^{3+}$ valency of this superlattice was further     confirmed by comparison with the bulk LaNiO$_3$. By  direct comparison of the absorption spectra in Fig. 2, the sharp difference between   the Ni charge state  in  the [1/1]$_{N}$ SLs and  the bulk like Ni$^{3+}$ spectra can be seen.

A few important  observations are due. First, the [1/1]$_3$ sample with LNO grown directly  on the STO surface clearly exhibits  a characteristic doublet with a 0.3$e$V shift to lower energy,  which matches the Ni$^{2+}$ XAS spectra of the NiO standard at the L$_2$ edge. This implies that the initial growth results in a phase analogous to heavily oxygen deficient bulk LaNiO$_{3-x}$ \cite{Sanchez}, such as LaNiO$_{2.5}$ with a Ni valence of 2+. As \textit{N} is further increased, the Ni  valence for the SLs on STO  moves progressively towards Ni$^{3+}$  as noted by the difference between the 6-repeat and 20-repeat samples. This result  implies that growing  the polar SLs on a non-polar or weakly-polar substrate such as STO results in a massive chemical compensation and electronic reconstruction during the initial growth. In contrast, the XAS spectra for the same set of SLs grown on LAO show that  Ni valence is very close to 3+ as directly evident from XAS on the SLs with \textit{N}=6 and 20. To exclude the compensating  charge transfer between Ti of STO and Ni of LNO, we  compared the\ Ti L-edge spectrum in a TEY mode (not shown)  to the Ti spectrum taken on a  bare STO  substrate\cite{Kareev2}. The  measurement unambiguously  confirms that no significant change of Ti$^{4+}$ valence  takes place near the film-substrate interface, likely due to the  high oxygen pressure during the deposition.

Next we investigated a possibility to preserve the fragile electronic state of Ni$^{3+}$ ions by fabricating a series of SLs on a 2 u.c. LAO buffer layer deposited on  STO. It is clear from the data shown in Fig. 2 for the \textit{N}=3 SL with a LAO buffer that the buffer greatly aids in restoring  the electronic state of  Ni towards 3+.
\begin{figure}[t]\vspace{-0pt}
\includegraphics[width=8.5cm, height=7.6cm]{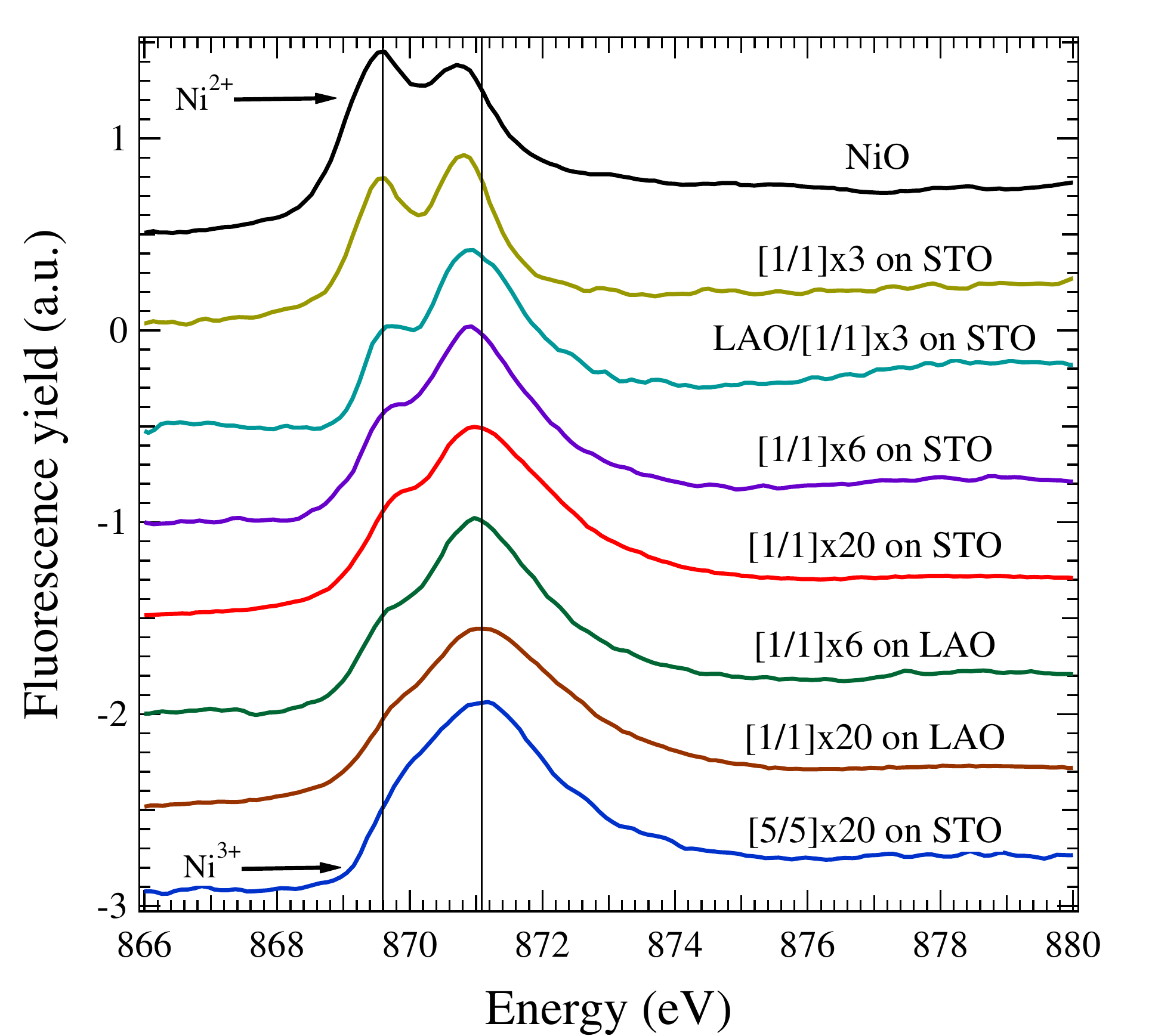}
\caption{\label{XAS} X-ray absorption spectra at the Ni L$_2$-edge of the [1/1]$_N$ superlattices  grown on STO and LAO. The top black curve represents the Ni$^{2+}$ standard from NiO.}
\end{figure}

Conductivity of bulk LNO has been shown to be sensitive to oxygen deficiency  which tightly correlates with Ni valence\cite{Sanchez}.
To clarify if  Ni$^{2+}$ is connected to oxygen deficiency, we annealed the SLs at 900$^o$C in flowing ultra-pure O$_2$   at 1 Atm  for 12 hours. Subsequent reduction in N$_2$  followed  by re-oxidation in O$_2$ was carried out under the same conditions. DC resistivity measurements were performed  after each  step using a 4-terminal geometry. This experiment  revealed  that prolonged  annealing in O$_2$ of the as-grown samples has no sizable effect on the resistivity. In contrast, the  reduction in nitrogen  had strongly increased the resistivity of the samples by several orders of magnitude, while the subsequent re-oxidation successfully restored it back to the as-grown level, indicating that the conductivity is not related to simple oxygen deficiency. This experiment provides strong evidence that the observed deviation of Ni valency from 3+ is not due to insufficient oxidation during the deposition but is likely  related to  the formation of acceptor-like complex defects caused by the polar discontinuity.


Fundamentally, the phase of a material deposited on the surface is determined by  enthalpy, $H$. While $H$ is controlled by the environmental thermodynamic parameters (\textit{e.g.} pressure and temperature), it also intimately depends on the atomic interactions on the surface (\textit{i.e}. wettability). Thus, the electrostatic interactions within a structure consisting of oppositely charged alternating atomic planes will result in a large increase to $H$ due to polar misfit\cite{hwang2}, and render the structure unstable. Experimentally, the probability to grow such a structure from the vapor phase is negligibly small. Instead, it has been proposed that such a polar mismatch can be generally resolved by transferring 0.5$e$ from  the surface to the polar-nonpolar interface\cite{Ohtomo, Hwang3}. However, the experimental observation of the large amount of Ni$^{2+}$ present in the ultra-thin SLs contradicts this model. To circumvent the issue,  we propose that the presence of Ni$^{2+}$ is likely associated with the polar compensation by the excess growth of an additional LaO plane. As illustrated in Fig. 4, at the initial stage of growth, this would result in a tri-layer structure like .../Ti$^{4+}$O$^{2-}_{2}$/La$^{3+}$O$^{2-}$/Ni$^{2+}$O$^{2-}_{2}$/La$^{3+}$O$^{2-}$, which contains a perovskite unit cell plus an extra LaO plane and remarkably possesses no total charge or dipole moment. Consequently, Ni is promoted to the 2+ charge state by the strong decrease of the electron potential due to the two LaO planes.

As the growth continues, the newly deposited [LaO/Ni(or Al)O$_2$]$_{m}$ ($m=0,1,2,...$) layers must be accommodated by adjusting the plane charge densities (the charge per u.c. in units of $|e|$) of the bottom NiO$_2$ plane and the surface plane. If the excess LaO plane remains on the surface, based on the charge neutrality and dipole compensation\cite{footnote}, one can show that its charge density equates to $(m+1)/(2m+1)$, while the Ni valence in the bottom NiO$_2$ plane becomes $3 -(m+1)/(2m+1)$. Note, setting $m$ to zero will correspond to the initial tri-layer structure.  On the other hand, if the terminating plane is Ni(Al)O$_2$, one can deduce that its charge density equates to $-m/2(m+1)$  with Ni valence in the bottom NiO$_2$ plane being $3-(m+2)/2(m+1)$ (here $m=0$ will correspond to a NiO$_{2}$ plane deposited on the top of the tri-layer, and each of the following LaO/Ni(Al)O$_{2}$ unit cells will increase $m$ by one). In either case, when $m \rightarrow \infty$, the charge density of the bottom NiO$_2$ plane changes from -2 to -1.5, corresponding to the evolution of the Ni valence from 2+ to 2.5+ \cite{footnote1}. This model lends theoretical support for the observation that the valence of the \textit{N}=6 SL is gradually approaching that of the \textit{N}=20 with no sign of a static Ni$^{2+}$ component \cite{footnote2}.
\begin{figure}[t]\vspace{-0pt}
\includegraphics[width=4.41cm, height=2.17cm]{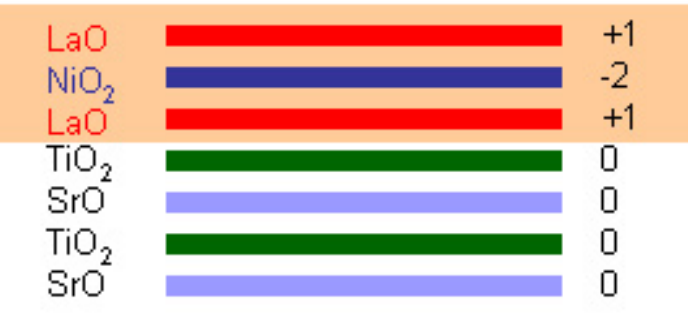}
\caption{\label{trilayer} The model tri-layer structure consisting of three atomic planes (charge densities on the right) resulted from the excess LaO growth at the initial stage of deposition on STO.}
\end{figure}

In practice, other mechanisms can certainly set in complicating the situation. In particular, non-stoichiometry such as atomic vacancies may play an important role. For instance, the electrostatic potential jump may decrease the enthalpy of formation of La vacancies in LaO planes and put related oxygen vacancies onto Ni(Al)O$_{2}$ planes  to reduce the dipole moment. Notice, the local field from the atomic vacancies is also capable of creating Ni$^{2+}$ due to a small charge transfer energy of the Ni$^{2+}$O$^{-}$ exciton.
 The precise correlation between the polarity problem and compensation by atomic defects requires further investigation.

%
In summary, by  monitoring  the Ni charge state as the control parameter, we have investigated the effect of polar mismatch on the electronic and structural properties of [1uc LNO/1uc LAO]$_N$ SLs.  The initial growth stage is found to experience a structural reconstruction accompanied by marked changes of the Ni electronic structure. We suggest that to a large extent these results can be explained by the new  tri-layer growth model, which corroborates the experimental data. Growing  LAO buffer on STO is found to efficiently circumvent the polar mismatch issue and allows  to preserve the electronically active Ni ions from changing the valence. These important  findings  are relevant for a wide range of complex-oxide materials  and should pave a way to growth of novel ultra-thin heterostructures with  controlled  electronic state of transition metal ions.

J.C. was supported by DOD-ARO under the Contract
No. 0402-17291 and NSF Contract No. DMR-0747808. Work at the Advanced Photon
Source, Argonne is supported by the U.S. Department of Energy, Office of
Science under Contract No. DEAC02-06CH11357.


\begin{thebibliography}{99}
\bibitem{footnote1}As $m \rightarrow \infty$, transfering an infinitizimally small charge $e/m$ by an infinitely large distance $mc$, where $c$ is the average c-axis lattice parameter, results in a finite dipole moment $ec$. This is why, at large thickness, both terminations provide the same result for the dipole moment, as any surface related dipole moment can be easily screened by transfering a small charge over the SL.
\bibitem{footnote2}In reality, a more gradual distribution may be expected for these charges that stems from the gradient contribution to $H$. The corresponding charge distribution function should minimize $H$ at each given thickness.
\end{thebibliography}
\end{document}